\documentclass[
    a4paper
    ,pra
    ,aps
    ,twocolumn
    ,nobibnotes
    ,notitlepage
]{revtex4-1}

\usepackage{amsmath,amssymb,geometry}

\usepackage{ifluatex,ifxetex}
\ifnum 0\ifxetex 1\fi\ifluatex 1\fi>0
    \usepackage[english]{babel}

    \usepackage{fontspec}
    \usepackage[math-style=ISO,bold-style=ISO]{unicode-math}
\else
    \ifpdf
        \usepackage{cmap}
    \fi
    \usepackage[T1]{fontenc}
    \usepackage[utf8]{inputenc}
    \usepackage[english]{babel}

\fi

\usepackage{graphicx}\graphicspath{{Graph/}{../Graph/}}

\usepackage{xcolor}
\definecolor{darkblue}{cmyk}{1.00, 0.50, 0.00, 0.40}
\definecolor{darkblue}{rgb}{0,0,.6}

\usepackage[
    colorlinks
        ,linkcolor=violet
        ,citecolor=teal
    ,unicode
    ,pdfpagemode=UseOutlines
    ,pdfdisplaydoctitle=true
    ,pagebackref=false
    ,pdfpagetransition=Wipe
    ,pdfusetitle
]{hyperref}

    \DeclareMathOperator{\e}{e}

    \DeclareMathOperator{\Rot}{rot}

    \newcommand{\re}{\mathrm{Re}}
    \newcommand{\im}{\mathrm{Im}}

    \newcommand{\parder}[2]{\frac{\partial #1}{\partial #2}}

    \newcommand{\dif}[1][]{\mathop{}\!\mathrm{d}
      \if\relax\detokenize{#1}\relax
      \else
        ^{\mkern-1.mu#1}\mkern-2.5mu 
      \fi}

    \renewcommand{\vec}[1]{\mathbf{#1}}

    \usepackage{xcolor}
    \definecolor{darkblue}{cmyk}{1.00, 0.50, 0.00, 0.40}

\begin{document}

\title{Dispersion Relation of a Surface Wave at a Rough Metal-Air Interface}

\author{Igor Kotelnikov}
\affiliation{Budker Institute of Nuclear Physics SB RAS}
\affiliation{Novosibirsk State University}

\author{Gennady Stupakov}
\affiliation{SLAC National Accelerator Laboratory,
Menlo Park, CA 94025, USA}

\begin{abstract}

    We derived a dispersion relation of a surface wave at a rough metal-air interface. In contrast to previous publications, we assumed that an intrinsic surface impedance due to a finite electric conductivity of the metal can be of the same order as the roughness-induced impedance. We then applied our results to the analysis of a long-standing problem of the discrepancy between the experimental data on the propagation of surface waves in the terahertz range of frequencies and the classical Drude theory.

\end{abstract}


\date{\today}

\maketitle

\section{Introduction}

The subject of Surface Waves (SW) propagating on a rough metal surface has attracted attention of many researches. There exists a vast literature devoted to this phenomenon. One of the earliest results was obtained by S. Rice  in 1951 \cite{Rice1951CPA_4_1097} who derived a dispersion relation for SW on a rough metal-air interface for the case of a metal with infinite conductivity (i.e. zero resistance). Although SW on a plane metal-air interface can, in theory, exist only if the metal possesses a finite electric resistivity, Rice has shown that roughness of the metal-air interface, in certain sense, replaces the electrical resistance so that SW on a rough surface can propagate even if the electrical resistance is negligible. Relatively recent reviews of more than 20 methods employed in solving this kind of problems can be found in \cite{Elfouhaily2004WRM_14_R1, Voronovich2013Wave}. In a form most relevant to the study of SW, important results are obtained in \cite{Mills1975PhysRevB_12_4036} and cited in \cite[p.~36]{Raether1988}.

In contrast to the earlier studies, in this paper, we consider SW taking into account a finite electrical resistance of the metal assuming that its effect in the SW dispersion is of the same order of magnitude as the surface roughness. The idea of our calculations is taken from Ref. \cite{BaneStupakov2000Linac_92} devoted to the beam wake field in an accelerator vacuum chamber caused by the wall roughness.  By comparing our results to Ref. \cite{Mills1975PhysRevB_12_4036} we conclude that the analysis in that article refers to the case where the effect of roughness is small compared to  the resistivity.

In a number of publications, the authors  start from a general treatment of scattering and absorption of electromagnetic waves on a rough boundary between air and a dielectric media with given permittivity $\varepsilon (\omega )$. In this paper, we employ a different approach based on the concept of a surface impedance. Note that this approach was successfully used earlier in our study of  SW on a conducting cylinder \cite{KotelnikovStupakov2015PLA_379_1187}. It greatly simplifies calculations by eliminating the need to computate electromagnetic fields inside the metal.

Below, we adhere to the following plan of presentation.

In section \ref{2.2} we remind key facts about dimensionless surface impedance and SW on plane metal-air interface. In section \ref{2.3}, we derive dispersion relation of SW on a sinusoidally corrugated surface for 1D case. In section \ref{2.4}, we extend this result to 2D corrugation. In section \ref{2.5}, we compute effective surface impedance for a rough surface. Finally, in Section \ref{2.6} we compare our theory with available experimental data.

\section{Surface wave at a flat metal-air interface}\label{2.2}

Consider a $p$-polarized wave that propagates in the $z$ direction along a plane metal-air interface. The magnetic field
    \begin{equation}
    \label{2:01}
    \vec{H}
    =
    \left\{H_{0},0,0\right\}
    \e^{ik_{z}z-\varkappa_{y}y-i\omega t}
    \end{equation}
of the wave in the upper half-space $z>0$ is characterized by the frequency $\omega $ and wavenumber $k_{z}$, as to  $\varkappa_{y}$, it can be found from the equation
    \begin{equation}
    \label{2:02}
    k_{z}^{2}-\varkappa _{y}^{2}=\omega ^{2}/c^{2}
    \end{equation}
and its real part should be positive for the wave to be considered as SW. As a standard theory of surface waves predicts (see e.g. \cite{Raether1988, KotelnikovStupakov2015PLA_379_1187}),  the parameters $k_{z}$ and $\varkappa_{y}$ for a SW propagating on a plane  metal-air interface are given by
    \begin{equation}
    \label{2:03}
    k_{z}
    =
    k \sqrt{\frac{\varepsilon }{1+ \varepsilon }}
    ,
    \qquad
    \varkappa_{y}
    =
    k \sqrt{-\frac{1}{1+ \varepsilon }}
    ,
    \end{equation}
where $\varepsilon=\varepsilon (\omega) $ is the permittivity of the metal, and $k=\omega /c$. The most simple model of a metal assumes that $\varepsilon = 1 - \omega _{p}^{2}/\omega^{2}$, where $\omega _{p}$ is called the plasma frequency. In such a model, $\varkappa_{y}$ is real (i.e. SW exist) if $\varepsilon <-1$, i.e. $\omega <\omega _{p}/\sqrt{2}$. An alternative description of metals adopts that $\varepsilon = 1 + 4\pi i \sigma/\omega $ with $\sigma $ being the electric conductivity. The latter model is more reliable for a limit of relatively low frequencies (e.g., terahertz, infrared and lower) where $\omega \ll |\sigma| $ and $|\varepsilon |\gg 1$. Then  Eq.~\eqref{2:03} can be approximated by
    \begin{equation}
    \label{2:04}
    k_{z}
    \approx
    k \left(
        1 - \frac{\xi^{2}}{2}
    \right)
    ,
    \qquad
    \varkappa_{y}
    \approx
    k i \xi
    ,
    \end{equation}
where
    \begin{equation}
    \label{2:05}
    \xi
    =
    \frac{1}{\sqrt{\varepsilon}}
    =
    (1-i)\sqrt{\frac{\omega}{8\pi\sigma}}
    \end{equation}
is the dimensionless surface impedance. On a rough surface, the dispersion relation for SW given by Eqs.~\eqref{2:03} and~\eqref{2:04} changes. We note however that Eq.~\eqref{2:03} can be kept by renormalizing the surface impedance $\xi\to\bar{\xi}$ so that one can say that the roughness changes the surface impedance.

Energy flux $\vec{S}=(c/8\pi)\re[\vec{E}\times \vec{H}^{\ast}]$ in SW is mainly directed along the metal-air interface and partially towards the metal surface. By designating the real and imaginary parts of $k_{z}=k_{z}'+ik_{z}''$ and $\varkappa_{y}=\varkappa_{y}'+i\varkappa_{y}''$ with the prime and double primes respectively, one can write
    \begin{equation}
    \label{2:06}
        \vec{S} = \left\{
        0
        , - \frac{\varkappa_{y}''}{k}
        , \frac{k_{z}'}{k}
        \right\}
        \frac{c|H_{0}|^{2}}{8\pi}
        \e^{-2(k_{z}''z+\varkappa_{y}'y)}
        .
    \end{equation}
By order of magnitude
    \begin{gather}
    \label{2:07}
    S_{y} = \mathcal{O}(\xi^{1})
    ,
    \qquad
    S_{z} = \mathcal{O}(\xi^{0})
    .
    \end{gather}
The energy flux $S_{z}$ in the direction of SW propagation is subject to the equation of the energy balance
    \begin{equation}
    \label{2:08}
    \parder{}{z}\int_{0}^{\infty} S_{z}\dif{y}
    =
    S_{y}\Bigr|_{y=0}
    .
    \end{equation}
It means that the energy density of SW decreases because of absorption in the metal and leads to easily verified relation
    \begin{equation*}
    -k_{z}'k_{z}''/\varkappa_{y}' = -\varkappa_{y}''
    \end{equation*}
which is a sequence of Eq.~\eqref{2:02}. Since $\varkappa_{y}'\sim\varkappa_{y}''=\mathcal{O}(\xi^{1})$ and $k_{z}'=\mathcal{O}(\xi^{0})$, this relation implies that $k_{z}''=\mathcal{O}(\xi^{2})$ in accord with Eq.~\eqref{2:04}. It is also worth noting that
    \begin{gather}
    \label{2:12}
    \int_{0}^{\infty }S_{z}\dif{y} = \mathcal{O}(\xi^{-1})
    .
    \end{gather}
The ordering \eqref{2:07} and \eqref{2:12} remains valid for SW on rough metal-air interface with the substitution $\xi\to\bar\xi$.

\section{1D corrugation}\label{2.3}

To get an idea of the effect of the roughness, we first consider a case of 1D surface corrugation assuming that the elevation of the metal-air interface is given by equation
    \begin{equation}
    \label{2.1:01}
    y = \mu h\sin(qz)
    ,
    \end{equation}
where $h$ and $q$ stand for the amplitude and wave number of the sinusoidal corrugation, and $\mu$ is a formal dimensionless parameter used below to distinguish between different orders of expansion over small amplitude $h$. The unit vector normal to the interface is given by
    \begin{equation}
    \label{2.1:02}
    \vec{n}
    =
    \frac{\left\{0,1,-{ \mu q h \cos (q z)}\right\}
    }{\sqrt{1 + h^2 \mu ^2 q^2 \cos ^2(q z)}}
    \end{equation}
and the corrugation is supposed to be shallow, i.e. its amplitude $h$ is much smaller than the period,  $qh\ll 1$.

Having in mind properties of SW outlined in Section \ref{2.2}, we will seek the magnetic field in the form of $p$-polarized wave
    \begin{equation}
    \label{2:01}
    \vec{H}= \left\{H_{x},0,0\right\}
    \end{equation}
as the sum
    \begin{equation}
    \label{2.2:04}
    H_{x}(y,x)
    =
    H_{0x}(y,x) + \delta H_{x}(y,z)
    \end{equation}
of a fundamental mode $H_{0x}(y,x)=H_{0}\e^{ik_{z}z-\varkappa_{y}y}$ with given amplitude $H_{0}$ and a satellite field $\delta H_{x}(y,z)$ that appears due to corrugation. Here and henceforth the time factor $\e^{-i\omega t}$ is dropped for the sake of brevity.
Recall that Eq.~\eqref{2:03} was derived for flat metal-air interface and should be changed on corrugated surface. Therefore we will consider $\varkappa_{y}$ as a free parameter to be found at the end of our calculations while $k_{z}$ is related to $\varkappa_{y}$ through the equation
    \begin{equation}
    \label{2:05}
    k_{z}=\sqrt{k^{2}+(\mu^{2}\varkappa_{y})^{2}}
    \end{equation}
instead of \eqref{2:03}. As will be shown below, the correction to $\varkappa_{y}$ due to surface corrugation is of second order in $\mu$ (i.e. in $h$). Therefore we assume that both $\varkappa_{y}$ and  $\xi$ are of second order in $\mu$ as we are most interested in analyzing the case where the effect of roughness is of order of the intrinsic surface impedance $\xi$ on its own. Thus, the fundamental harmonic in SW should be sought in the form
    \begin{equation}
    \label{2.2:04c}
    H_{x0}(y,z) = H_{0} \e^{i k_{z}z-\mu^{2}\varkappa_{y}y}
    .
    \end{equation}

As to the satellite field, we seek it in the form of two waves, exponentially decaying as $y$ rises:
    \begin{multline}
    \label{2.2:04d}
    \delta H_{x}(y,z)
    =
    \mu B_{+}\e^{i(k_{z}+q)z - \sqrt{(k_{z}+q)^{2}-k^{2}}y}
    +
    \\
    +
    \mu B_{-}\e^{i(k_{z}-q)z - \sqrt{(k_{z}-q)^{2}-k^{2}}y}
    .
    \end{multline}
The magnetic fields  \eqref{2.2:04c} and \eqref{2.2:04d} obey the Helmholtz equation
    \begin{equation}
    \label{2.2:05}
    \parder{^{2}H_{x}}{y^{2}}
    +
    \parder{^{2}H_{x}}{z^{2}}
    +
    k^{2}H_{x}
    =0.
    \end{equation}
Note however that the satellite waves with wavenumbers $k_{z}\pm q$ do not represent eigenmodes by itself (i.e. they are not a proper solution of the boundary value problem on corrugated surface) and, hence, they cannot exist without fundamental mode with the wavenumber $k_{z}$. Instead, the fundamental SW plus the satellite waves form a proper mode of the rough metal-air boundary.

The electric field is expressed through $\vec{H}$ by
    \begin{equation}
    \label{2.2:06}
    \vec{E} = \frac{i}{k}\Rot\vec{H}
    .
    \end{equation}
At the metal-air interface the tangential part of the electric field
    \begin{equation}
    \label{2.2:11}
    \vec{E}_{t}= - [\vec{n}\times [\vec{n}\times \vec{E}]]
    \end{equation}
is related to the magnetic field by the boundary condition \cite{Leontovich1944IANUSSR(eng)}
    \begin{equation}
    \label{2.2:12}
    \vec{E}_{t}
    =
    \mu^{2}
    \xi
    \left[\vec{n}\times \vec{H}\right]
    .
    \end{equation}

Our goal is to find a replacement for the dispersion relation \eqref{2:03} on a flat metal-air interface which would be valid on a rough surface. Due to Eq.~\eqref{2:05} this goal will be achieved if we compute $\varkappa_{y}$ up to the second order on $\mu$. To do that we put the expressions for the electric and magnetic fields in Eq.~\eqref{2.2:12}, expand the result into a series over parameter $\mu$ and separate terms with different dependency on $z$, i.e., the terms containing $\e^{ik_{z}z}$, $\e^{i (k_{z}z\pm q)z}$, $\e^{i (k_{z}z\pm 2q)z}$, e.t.c. This procedure yields a set of equations for unknown coefficients $B_{\pm}$ and $\varkappa _{y}$.

In zeroth order of expansion on $\mu$, we obtain only trivial equations since all terms in Eq.~\eqref{2.2:12} yield zero.

The first order of the expansion yields $2$ equations for the satellite amplitudes $B_{\pm}$ after separating terms with $\e^{i(k_{z}\pm q)z}$ factors. Solving these equations and noting that $k_{z}=k$ in this order gives
    \begin{equation}
    \label{2.3:23}
    B_{\pm}
    =
    -
    \frac{
        i k q h  H_{0}
    }{
        2 \sqrt{(k\pm q)^{2}-k^{2}}
    }
    .
    \end{equation}

Finally, second order of the series yields $3$ independent equations after separating terms proportional to $\e^{ik_{z}z}$, $\e^{i(k_{z}+2q)z}$, and $\e^{i(k_{z}-2q)z}$. The last two equations could allow determining amplitudes of the second order satellites with wavenumbers $k_{z}\pm 2q$ but we did not include them in Eq.~\eqref{2.2:04d}. And the former equation allows computing the wavenumber $\varkappa _{y}$. Noting that in this order  again  $k_{z}=k$ we find
    \begin{equation}
    \label{2.2:24}
    \varkappa _{y}
    =
        i k\zeta
        +
        \frac{h^2 k^{2} q^2}{4\sqrt{q (q-2k)}}
        +
        \frac{h^2 k^{2} q^2}{4\sqrt{q (2 k+q)}}
    .
    \end{equation}
The corrugation terms here, which are proportional to $h^{2}$, are additive to the intrinsic impedance $\xi$. Therefore one can use the dispersion relation \eqref{2:03} for SW on corrugated metal-air interface after substitution of $\xi$ for the effective surface impedance
    \begin{equation}
    \label{2.2:25}
    \bar{\xi}
    =
    \xi
    -
    \frac{ih^2 k q^2}{4\sqrt{q (q-2k)}}
    -
    \frac{ih^2 k q^2}{4\sqrt{q (2 k+q)}}
    .
    \end{equation}
In case $\xi=0$ Eq.~\eqref{2.2:24} coincides with Eq.~(6.10) in Ref.~\onlinecite{Rice1951CPA_4_1097}.

The square roots $\sqrt{q (q \pm 2 k)}$ in Eqs.~\eqref{2.2:24}  and~\eqref{2.2:25} originate from the $y$ components $\sqrt{(k_{z}\pm q)^{2} - k^{2}}$ of the wave vector of satellite waves. Therefore the sign of these roots, when  $q (q \pm 2 k) < 0$ and they are imaginary, should be chosen in such a way that an exponentially decaying satellite wave transforms into a wave freely propagating out of the metal. Hence,
    \begin{equation}
    \label{2.2:26}
    \sqrt{q (q \pm 2 k)} \to -i \sqrt{|q (q \pm 2 k)|}
    \end{equation}
if $q (q \pm 2 k) < 0$. It can be readily seen that of the two roots at a given value of $q$ only one is imaginary (an, hence, only one of the two satellite waves is freely propagating) if
    \begin{equation}
    \label{2.2:27}
    -2k < q < +2k
    \end{equation}
and that both roots are real if
    \begin{equation}
    \label{2.2:28}
    |q|> 2k
    \end{equation}
(and both satellite waves are decaying).

Without lost of generality we assume below in this Section that $q>0$ and  focus on the case $0<q<2k$. Then, the surface corrugation attracts  additional energy flux  in fundamental SW \emph{towards} the metal. Normal component of the energy flux at the metal-air interface (at $y=0$) is
    \begin{multline}
    \label{2.2:31}
    S_{y}\Bigr|_{y=0}
    = \frac{c}{8\pi}\re(E_{z}H_{x}^{\ast})
    = \frac{c}{8\pi}\re\left(
        \frac{i\varkappa _{y}}{k}|H_{0}|^{2}
    \right)
    =
    \\
    =
    - \xi'\frac{c|H_{0}|^{2}}{8\pi}
    - \frac{h^{2}k q^{2}}{4\sqrt{|q(q - 2k)|}}
        \frac{c|H_{0}|^{2}}{8\pi}
    \end{multline}
where $\xi'=\re(\xi)>0$. The second term in \eqref{2.2:31} describes the energy influx caused by the surface corrugation. However exactly the same energy flux is reradiated \emph{outwards} as satellite wave. Indeed, noting that a free propagating wave in the case $0<q<2k$ has the amplitude $B_{-}$ and radiates at the angle $\theta_{-}=\arcsin({\sqrt{|q(q - 2k)|}}/{k})$, we obtain
    \begin{multline}
    \label{2.2:32}
    S_{y}
    = \frac{c}{8\pi} |B_{-}|^{2} \sin\theta _{-}
    =
    \\
    = \frac{c}{8\pi} \left|
        \frac{ikqhH_{0}}{2\sqrt{q(q - 2k)}}
    \right|^{2}
    \frac{\sqrt{|q(q - 2k)|}}{k}
    =
    \\
    =
    \frac{c}{32\pi} \frac{h^{2}k q^{2}}{\sqrt{|q(q - 2k)|}}
    |H_{0}|^{2}
    .
    \end{multline}
This process can be categorized as a scattering of SW on the surface corrugation. It leads to additional weakening of primary SW according to Eq.~\eqref{2:08} just as if the scattered energy flux would be absorbed by the metal.

\section{2D corrugation}\label{2.4}

Assume now that a sinusoidal corrugation is not aligned with the direction of propagation of SW. Let the metal-air interface be given by the equation $F=0$, where
    \begin{gather}
    \label{4:01}
    F = y - \mu h\sin(\vec{q}\cdot \vec{x} + \psi)
\intertext{with an arbitrary 2D vector}
    \label{4:02}
    \vec{q}=(q_{x},0,q_{z})
    ,
\intertext{radius-vector}
    \label{4:03}
    \vec{x}=(x,0,z)
    ,
    \end{gather}
an arbitrary phase $\psi$, and the amplitude of corrugation $h$.  The unit vector normal to the interface that enters Eqs.~\eqref{2.2:06}, \eqref{2.2:11}, and \eqref{2.2:12} is now given by
    \begin{equation}
    \label{4:04}
    \vec{n}
    =
    \frac{\nabla F}{|\nabla F|}
    .
    \end{equation}

The fundamental SW is now sought in the form
    \begin{equation}
    \label{4:06}
    \vec{H}_{0}=\left(
        H_{0}, \mu^{2} B,0
    \right)\e^{ik_{z}z - \mu^{2}\varkappa_{y}y}
    ,
    \end{equation}
where $B$ is an unknown coefficient to be found; this form is justified by the final result.

The satellite waves are characterized by the wave vectors
    \begin{equation}
    \label{4:07}
    \vec{k}_{\pm}
    =
    \vec{k}\pm\vec{q}
    +
    \left(0,\sqrt{k^{2}-(\vec{k}\pm\vec{q})^{2}},0\right)
    ,
    \end{equation}
where
    \begin{equation}
    \label{4:08}
    \vec{k} = \{0,0,k_{z}\}
    ,
    \end{equation}
and
    \begin{equation}
    \label{4:12}
    |\vec{k}_{\pm}|=k
    .
    \end{equation}
Each  wave has $2$ independent polarizations. We generate polarization unit vectors $\vec{v}_{\pm}$ and $\vec{u}_{\pm}$ using the following procedure. First we select an initial vector $\vec{v}_{0}$, say
    \begin{equation}
    \label{4:14}
    \vec{v}_{0}=(1,0,0)
    \end{equation}
or $\vec{v}_{0}=(0,1,0)$ or $\vec{v}_{0}=(0,0,1)$. Then we compose a vector perpendicular to $k_{+}$:
    \begin{equation}
    \vec{v}_{1}=\vec{v}_{0} - \left(\vec{v}_{0}\cdot \vec{k_{+}}\right)
    \vec{k_{+}}/k^{2}
    \end{equation}
and normalize it:
    \begin{equation}
    \label{4:16}
    \vec{v}_{+}=\vec{v}_{1}/|\vec{v}_{1}|
    .
    \end{equation}
This yields the first unit vector $\vec{v}_{+}$ perpendicular to $\vec{k}_{+}$. Rotating it by $90^{\circ}$ about the direction of the wave vector $\vec{k}_{+}$ yields the second unit vector:
    \begin{equation}
    \label{4:17}
    \vec{u}_{+}=[\vec{v}_{+}\times\vec{k}_{+}]/k
    .
    \end{equation}
To compose similar vectors $\vec{v}_{-}$ and  $\vec{u}_{-}$ for the second satellite wave it is sufficient to perform the substitution $\vec{q}\to-\vec{q}$ in $\vec{v}_{+}$ and  $\vec{u}_{+}$. Taking different initial vectors $\vec{v}_{0}$ generates different sets of polarization vectors. We chose a set that originates from Eq.~\eqref{4:14}:
    \begin{widetext}
    \begin{equation}
    \begin{aligned}
    \label{4:18}
    \vec{v}_{\pm}
    &=
    \left\{
        \frac{\sqrt{k^2-q_{x}^2}}{k}
        ,
        \mp
        \frac{
            q_{x}\sqrt{k^{2}-q_{x}^2-(k_{z}\pm q_{z})^{2}}
        }{
            k\sqrt{k^2-q_{x}^2}
        }
        ,
        -\frac{
            q_{x} (q_{z} \pm k)
        }{k\sqrt{k^2-q_{x}^2}}
    \right\}
    ,
    \\
    \vec{u}_{\pm}
    &=
    \left\{
        0
        ,
        -
        \frac{k \pm q_{z}}{\sqrt{k^2-q_{x}^2}}
        ,
        \frac{
            \sqrt{k^{2}-q_{x}^2-(k_{z}\pm q_{z})^{2}}
        }{
            \sqrt{k^2-q_{x}^2}
        }
    \right\}
    .
    \end{aligned}
    \end{equation}
Now a satellite wave can be written as
    \begin{equation}
    \label{4:21}
    \delta \vec{H}
    =
    \mu\left(
        V_{+}\vec{v}_{+}
        +
        U_{+}\vec{u}_{+}
    \right)
    \e^{i\vec{k}_{+}\cdot \vec{x}}
    +
    \mu\left(
        V_{-}\vec{v}_{-}
        +
        U_{-}\vec{u}_{-}
    \right)
    \e^{i\vec{k}_{-}\cdot \vec{x}}
    .
    \end{equation}
Repeating the procedure described in the previous Section and again noting that $k_{z}=k$ within the desired accuracy, we find
    \begin{gather}
    \label{4:22}
    V_{\pm}
    =
    \frac{
         k q_{z} \pm q_{x}^2
    }{
        2 \sqrt{k^2-q_{x}^2}
        \sqrt{k^{2}-q_{x}^2-(k\pm q_{z})^{2}}
    }\,
    \e^{\pm i\psi} k h H_{0}
   ,
   \\
    \label{4:23}
    U_{\pm}
    =
    -\frac{ q_{x}}{2 \sqrt{k^2-q_{x}^2}}\,
    \e^{\pm i\psi}k h H_{0}
    ,
    \\
    \label{4:24}
    \varkappa_{y}
    =
    ik\xi
    +
    \frac{i}{4} q_{z}^2 k^{2} h^2
    \left(
        \frac{1}{\sqrt{k^{2}-q_{x}^2-(k+q_{z})^{2}}}
        +
        \frac{1}{\sqrt{k^{2}-q_{x}^2-(k-q_{z})^{2}}}
    \right)
    ,
    \\
    \label{4:25}
    B
    =
    -
    \frac{1}{4} q_{z}q_{x} k h^2 H_{0}
    \left(
        \frac{1}{\sqrt{k^{2}-q_{x}^2-(k+q_{z})^{2}}}
        +
        \frac{1}{\sqrt{k^{2}-q_{x}^2-(k-q_{z})^{2}}}
    \right)
    .
    \end{gather}
    \end{widetext}
It has been checked that the specific expressions for the coefficients $\varkappa_{y}$ and $B$ are not sensitive to the choice of polarization vectors as well as the expression for the vectors
    \begin{multline}
    \label{4:26}
    \vec{B}_{\pm}
    \equiv
    V_{\pm}\vec{v}_{\pm} + U_{\pm}\vec{u}_{\pm}
    =
    \\
    =
    \left\{
        \frac{h
            \left(k q_{z}\pm q_{x}^2\right)H_{0}\e^{\pm i\psi}
        }{
            2 \sqrt{k^{2}- q_{x}^2 - (q_{z}\pm k)^{2}}
        }
        ,
        \frac{q_{x} h \,H_{0} \e^{\pm i\psi}}{2}
        ,
    \right.
    \\
    \left.
        \pm
        \frac{
            q_{x} q_{z} h \,H_{0}\e^{\pm i\psi}
        }{
            2 \sqrt{k^{2}- q_{x}^2 - (q_{z}\pm k)^{2}}
        }
    \right\}
    .
    \end{multline}
The meaning of the coefficients $\varkappa_{y}$ and $B$ can be deduced from the expressions for the field of fundamental harmonic at the plane $y=0$:
    \begin{equation*}
    \vec{E} = \{B,-H_{0}, (i\varkappa_{y}/k)H_{0} \}
    ,
    \quad
    \vec{H} = \{H_{0},B, 0 \}
    .
    \end{equation*}
One can see that the coefficient $B$ stands for additional energy flux along the metal-air boundary since
    \begin{multline*}
    S_{z}
    =
    \frac{c}{8\pi} \re
    \left(
        E_{x}H_{y}^{*}-E_{y}H_{x}^{*}
    \right)
    =
    \\
    =
    \frac{c}{8\pi} \left(
        |H_{0}|^{2}+ |B|^{2}
    \right)
    ,
    \end{multline*}
however the addition of $cB^{2}/8\pi$ exceeds the accuracy of our calculations. The coefficient $\varkappa_{y}$ is responsible for the energy flux in the direction towards the metal:
    \begin{multline}
    \label{4:33}
    S_{y} = \frac{c}{8\pi} \re\left(
        -E_{x}H_{z}^{*} + E_{z}H_{x}^{*}
    \right)
    =
    \\
    =
    \frac{\re(i\varkappa_{y})}{k}\frac{c|H_{0}|^{2}}{8\pi}
    .
    \end{multline}

Further analysis follows that of Section \ref{2.3}. One can show that the additional roughness-induced flux is directed towards the metal and appears only if any of the inequalities
    \begin{equation}
    \label{4:35}
    (q_{z}\pm k)^{2} + q_{x}^{2} < k^{2}
    \end{equation}
holds. This flux is re-radiated in the form of a freely propagating satellite wave.

\begin{figure}
  \centering
  \includegraphics[width=0.5\textwidth]{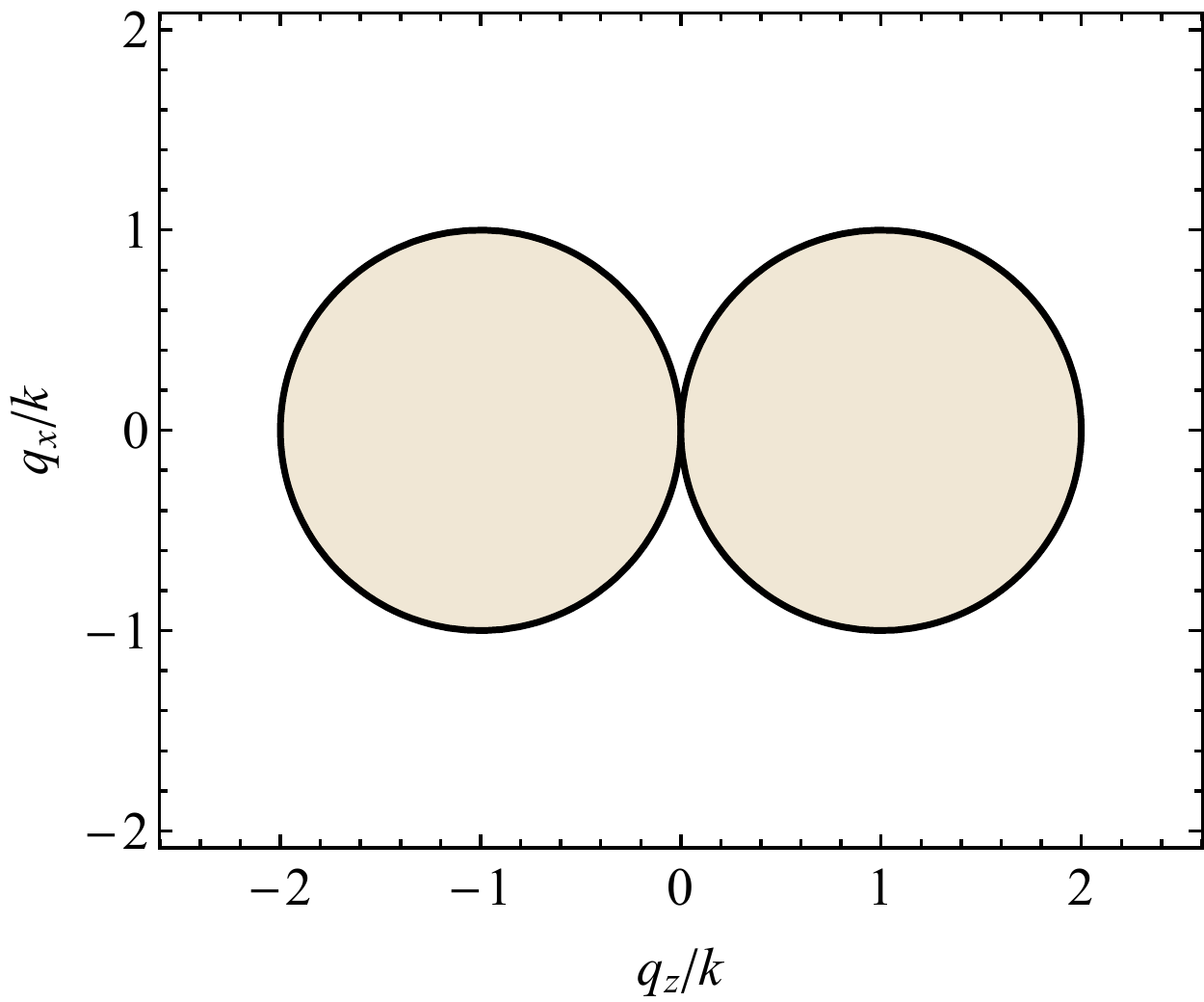}
  \caption{Radiation Zone}\label{fig:RadiationZone}
\end{figure}
As shown in Fig.~\ref{fig:RadiationZone}, the radiation zone \eqref{4:35} includes the interiors of two adjacent circles in the plane of vector $\vec{q}$. As the circles have no common parts except for the point $\vec{q}=0$, only one of the two summands in Eq.~\eqref{4:24} can contribute to the radiation for a given $\vec{q}$. The re-radiated energy flux is given
    \begin{equation}
    \label{4:41}
    S_{y} =
    \frac{k q_{z}^{2} h^{2}}{4\sqrt{k^{2}-q_{x}^{2}-(q_{z} \pm k)^{2}}}
    \frac{c|H_{0}|^{2}}{8\pi}
    ,
    \end{equation}
where of the two signs in the denominator should be selected such that obeys the condition \eqref{4:35}.

Singularity in Eq.~\eqref{4:41} at
    \begin{equation}
    \label{4:42}
    (q_{z}\pm k)^{2} + q_{x}^{2} = k^{2}
    \end{equation}
corresponds to scattering of primarily SW into SWs of different direction but without change of the absolute value of the wave vector (see Ref. \onlinecite[p.~36]{Raether1988}). Such SWs represent eigenmodes which can exist by themselves without bound to original SW. In theory of Ref.~\cite{Mills1975PhysRevB_12_4036} the contribution of such SWs is computed as a residue in a complex plane and gives a negligible correction. In our theory, this contribution is zero since the singularity in Eq.~\eqref{4:41} is integrable (see next Section).


\section{Impedance of a rough surface}\label{2.5}

With a small modification, our analysis can be also applied to the case of a surface that has a random roughness profile. A realistic metal-air interface can be modeled by a mixture of corrugations with different vectors $\vec{q}$:
    \begin{equation}
    \label{5:01}
    y(\vec{x}) = \sum_{\vec{q}} f(\vec{q})\e^{i\vec{q}\cdot \vec{x}}
    \to
    \int \frac{\dif[2]q}{(2\pi)^{2}}\,f(\vec{q})\,\e^{i\vec{q}\cdot \vec{x}}
    .
    \end{equation}
Since $y$ is a real function, the coefficients $f(\vec{q})$ satisfy
    \begin{equation}
    \label{5:02}
    f(\vec{q}) = f^{\ast}(-\vec{q})
    .
    \end{equation}
In terms of previous Section,
    \begin{equation*}
    f(\vec{q}')
    =
    \frac{h}{2i}
    \e^{i\psi}\delta (\vec{q}'-\vec{q})
    -
    \frac{h}{2i}
    \e^{-i\psi}\delta (\vec{q}'+\vec{q})
    .
    \end{equation*}
It is usually assumed that an average (in a certain sense)  value of $y(\vec{x})$ is zero,
    \begin{equation}
    \label{5:04}
    \langle y(\vec{x})\rangle = 0
    ,
    \end{equation}
and the correlation function $\langle y(\vec{x})y(\vec{x}')\rangle$ depends only on the difference $\vec{x}-\vec{x}'$:
    \begin{equation}
    \label{5:05}
    \langle y(\vec{x})y(\vec{x}')\rangle
    \equiv
    W(\vec{x}-\vec{x}')
    .
    \end{equation}
Averaging can be understood either as averaging over the stochastic phases $\psi$ or averaging over the coordinate $\vec{x}+\vec{x}'$ under the assumption that the stochastic properties of the metal-air interface are uniform. In terms of previous Section,
    \begin{equation*}
    W(\vec{x})
    =
    \frac{1}{2}\,h^{2}\cos(\vec{q}\cdot \vec{x})
    .
    \end{equation*}

Putting the integral \eqref{5:01} in Eq.~\eqref{5:05} leads to the conclusion that
    \begin{equation}
    \label{5:07}
    \langle f(\vec{q})\,f(\vec{q}')\rangle
    =
    (2\pi)^{2} G(\vec{q})\,\delta(\vec{q}+\vec{q}')
    ,
    \end{equation}
where
    \begin{equation}
    \label{5:08}
    G(\vec{q})
    =
    \int \dif[2]x\, W(\vec{x})\,\e^{-i\vec{q}\cdot \vec{x}}
    .
    \end{equation}
In terms of the previous Section,
    \begin{equation*}
    G(\vec{q}')
    =
    \frac{h^{2}}{4}\,
    (2\pi)^{2}
    \left[
        \delta(\vec{q}'+\vec{q})
        +
        \delta(\vec{q}'-\vec{q})
    \right]
    .
    \end{equation*}
Hence, to generalize the already known results for the case of a rough surface it is sufficient to perform the substitution
    \begin{equation}
    \label{5:11}
    \frac{h^{2}}{4}\left(\ldots\right)
    \to
    \int \frac{\dif[2]q}{(2\pi)^{2}} G(\vec{q})
    \left(\ldots\right)
    \end{equation}
in Eq.~\eqref{4:24}. The rule \eqref{5:11} yields the final expression for the effective surface impedance:
    \begin{equation}
    \label{5:12}
    \bar\xi
    =
    \xi
    +
    \int \frac{\dif[2]q}{(2\pi)^{2}}
    \frac{
        q_{z}^2 k\,G(\vec{q})
    }{
        \sqrt{k^{2}-q_{x}^2-(k-q_{z})^{2}}
    }
    .
    \end{equation}
Recall that
    \begin{equation*}
    \sqrt{k^{2}-q_{x}^2-(k-q_{z})^{2}}
    =
    i\sqrt{|k^{2}-q_{x}^2-(k-q_{z})^{2}|}
    \end{equation*}
if $k^{2}<q_{x}^2+(k-q_{z})^{2}$.

\begin{figure}
  \centering
  \includegraphics[width=\columnwidth]{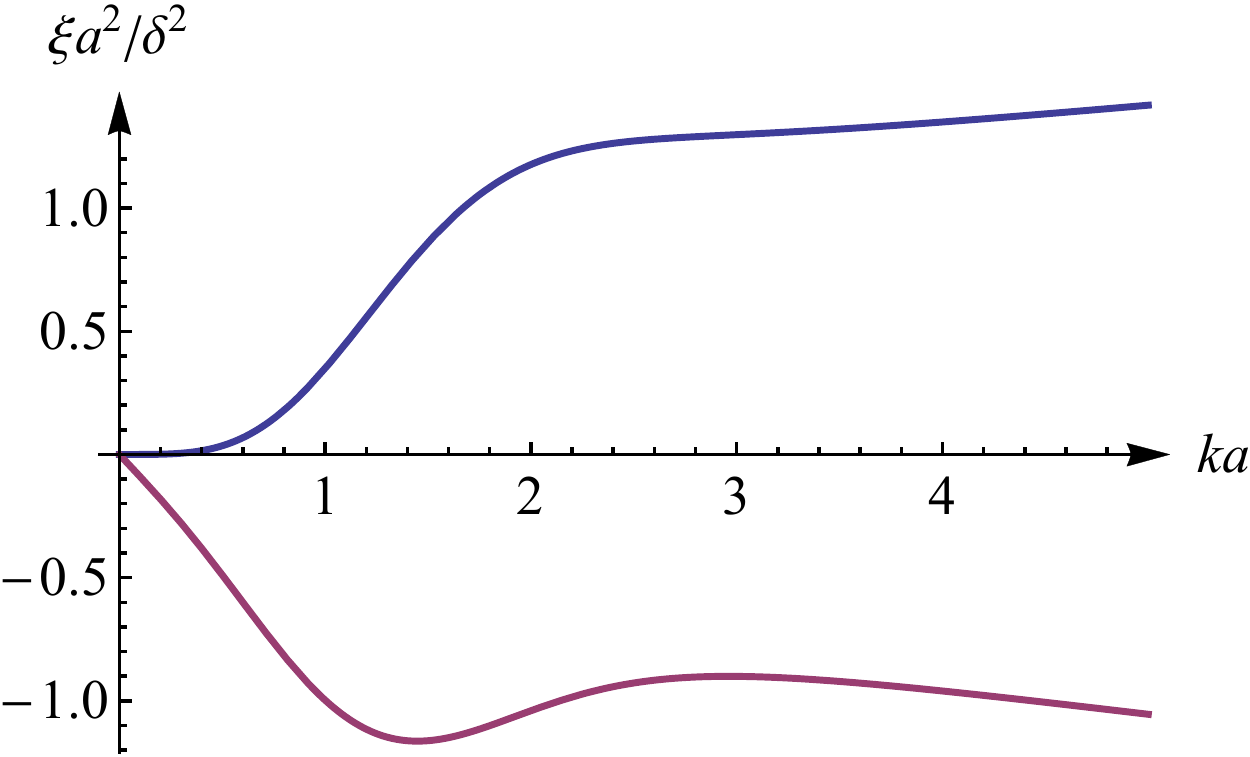}
  \caption{
    (Color online)
    An additional surface impedance caused by surface roughness for a Gaussian correlation function: blue (upper curve) is the real part of $\bar\xi$, purple (bottom curve) is the imaginary part. }\label{fig:SurfaceImpedance}
\end{figure}

A popular model for the correlation function is Gaussian:
    \begin{equation}
    \label{5:16}
    W(\vec{x}) = \delta ^{2} \e^{-|\vec{x}|^{2}/a^{2}}
    ,
    \end{equation}
where $\delta $ is the r.m.s.\ height of the roughness and $a$ is an average radius of the roughness bumps. Then
    \begin{equation}
    \label{5:17}
    G(\vec{q}) =
    \pi \delta ^{2} a^{2}\e^{-|\vec{q}|^{2}a^{2}/4}
    .
    \end{equation}
A numerically computed surface impedance is shown in Fig.~\ref{fig:SurfaceImpedance}.

For the most interesting case of small-size bumps, $ka\ll1$, we have
    \begin{equation}
    \label{5:18}
    \bar\xi
    =
    \xi
    +
    \frac{\delta ^{2}}{a^{2}}
    \left[
        -
        i \frac{\sqrt{\pi}}{2} ka
        +
        \frac{2}{3} k^{4}a^{4}
    \right]
    .
    \end{equation}
Here the first term in the square brackets dominates but it seems that it is missed in earlier theories (see \cite{Mills1975PhysRevB_12_4036, Raether1988}).

\bigskip
\section{Discussion}\label{2.6}

The main theoretical results of this paper are given by Eqs.~\eqref{5:12} and~\eqref{5:18}. We have shown that the intrinsic surface impedance caused by the finite resistivity of metal and an additional roughness-induced impedance are additive when they are of the same order of magnitude. Previous studies \cite{Mills1975PhysRevB_12_4036,Raether1988} dealt with the case of either a small effect of the surface roughness or, on the contrary, with the case of zero intrinsic surface impedance \cite{Rice1951CPA_4_1097}. Finally, we concluded that the most important first term in the square brackets in Eq.~\eqref{5:18} was missed in earlier theories.

In Section~\ref{2.3} we have outlined a clear picture of energy flows in the surface wave on a rough boundary metal-air interface. In particular, we have shown that the roughness-induced energy flux in SW towards the metal-air interface is reradiated back at a slope angles provided that inequality~\eqref{4:35} holds, and otherwise no additional flux arises.

To compare our results with experimental data summarized in Refs.~\cite{Gerasimov+2011APL_98_171912, Pandey+2016AdvPhysX_1_176}, one needs to compute vertical and horizontal scale-lengths of a surface wave.

The vertical scale-length $L_{y}$ of SW can be found from equation
    \begin{equation}
    \label{2.2:34}
    \frac{1}{L_{y}}
    =
    \re(\varkappa_{y})
    =
    -k\,\im(\bar\xi)
    ,
    \end{equation}
and, as follows from Eq.~\eqref{5:18}, is strongly affected by the surface roughness. On the contrary, experimental data \cite{Gerasimov+2011APL_98_171912, Pandey+2016AdvPhysX_1_176} supports the conclusion that in the terahertz range of frequencies $L_{y}$ conforms the classical Drude theory which does not take into account the effect of surface roughness.


Attenuation of SW in the direction of its propagation appears in the 4th order in $\mu$ as can be seen from the first of Eq.~\eqref{2:04} with $\bar\xi$ taken instead of $\xi$:
    \begin{equation}
    \label{2.2:33}
    k_{z} = k\left(
        1 - \frac{\bar{\xi}^{2}}{2}
    \right)
    .
    \end{equation}
It gives the following expression for the horizontal scale $L_{z}$:
    \begin{equation}
    \label{2.2:35}
    \frac{1}{L_{z}}
    =
    -\frac{1}{2}k\,\im(\bar\xi^{2})
    =
    \frac{1}{L_{y}} \re(\bar\xi)
    .
    \end{equation}
Since the roughness-induced real part of the surface impedance is smaller than the imaginary part, Eq.~\eqref{2.2:35} means that $L_{z}$ should also conform the Drude theory. However this conclusion is in contradiction with the above cited experimental data.  Therefore, it does not seem that the experimentally observed reduction of $L_{z}$ by 2-3 orders of magnitude as compared to the Drude theory can be caused by radiation losses of SW energy as suggested in Ref.~\cite{Gerasimov+2011APL_98_171912, Pandey+2016AdvPhysX_1_176}. Nevertheless, we note that the actual correlation function of a rough surface may significantly differ from the Gaussian one of Eq.~\eqref{5:16}, which was used in our calculations. For example, the correlation function can be non-monotonic as reported in Ref.~\cite{Zhizhin+1985PZhTF_11_951(eng)}. This would mean an existence of a dominated wavenumber $\vec{q}$ in the power spectrum of the surface roughness. In turn, it could enhance the effect of the radiation losses of the SW power, especially if $\vec{q}$ is in proximity of solid circles in Fig.~\ref{fig:RadiationZone}.

In our opinion, it is feasible that the above mentioned experimental results might be attributed to the effect of enhanced ohmic losses in thin metal films as it is also discussed in Ref.~\cite{Gerasimov+2011APL_98_171912, Pandey+2016AdvPhysX_1_176}.

\section{Acknowledgements}

We are grateful to V.~Gerasimov and B.~Knyazev who attracted our attention to the problem discussed in this paper.

The work by I. Kotelnikov was supported by Russian Science Foundation (project N 14-50-00080).

The work by G. Stupakov was supported by the Department of Energy,
contract DE-AC03-76SF00515

%

%

\end{document}